\documentclass[a4paper,11pt]{article}
\usepackage{pos}

\usepackage{hyperref}

\hypersetup{colorlinks=true,linkcolor=blue,urlcolor=blue}
\title{The four-gluon vertex from lattice QCD}

\author[]{Manuel Colaço}
\author*[]{Orlando Oliveira}
\author[]{Paulo J Silva}

\affiliation[]{CFisUC. Department of Physics, University of Coimbra, 3004-516 Coimbra, Portugal}

\emailAdd{mailto:mancol@amu.edu.pl}
\emailAdd{orlando@uc.pt}
\emailAdd{psilva@uc.pt}

\abstract{The four-gluon one-particle irreducible Green function contributes to various quantities with phenomenological relevance. An example where the four-gluon plays a role is the determination of the gluon propagator, a basic building block for QCD, using continuum methods. This four leg Green function is poorly known and we are only starting to grasp its non-perturbative structure. Here, we report on the computation of the one-particle irreducible four-gluon Green function, in the Landau gauge, with lattice simulations. Besides stating the problems associated with the computation, several form factors that characterise this Green function are measured.}

\FullConference{The 41st International Symposium on Lattice Field Theory (LATTICE2024)\\
 28 July - 3 August 2024\\
Liverpool, UK\\}

\begin{document}
\maketitle

\section{Introduction and Motivation}

The classical Lagrangian density of Yang-Mills theory has a quartic term in the gauge field and, therefore, in a perturbative solution of the corresponding quantum field theory
there is a non-vanishing tree level Feynman rule associated with the Green function with four gauge field legs. In QCD, the corresponding four-gluon one-particle irreducible (1PI)
Green function appears, for example, in the Dyson-Schwinger equation for the gluon propagator, in a two-loop diagram, see e.g. \cite{Eichmann:2021zuv}. 
Certainly, the four-gluon 1PI plays a role, even if it is not the dominant contribution, in the dynamics of gluons, see e.g. 
\cite{Driesen:1998xc,Ewerz:1999gk,Dumitru:2010mv,Abreu:2017xsl}. 
Moreover, it is possible to define a renormalization group invariant QCD charge from this vertex \cite{Kellermann:2008iw}.
These are among the reasons why it is important to know the structure of this Green function, specially in the non-perturbative regime of the theory, 
and what are the dominant tensors that contribute to this 1PI function.

In the non-perturbative regime, the four-gluon 1PI Green function has been within the continuum formulation of QCD by several authors
\cite{Kellermann:2008iw,Binosi:2014kka,Cyrol:2014kca,Cyrol:2016tym,Cyrol:2017ewj,Huber:2018ned,Huber:2020keu,Aguilar:2024fen,Barrios:2024ixj,Aguilar:2024dlv}. 
Recently, a number of studies using lattice methods
investigated this Green function either in collinear kinematics \cite{Catumba:2021qbh,Colaco:2023qin,Colaco:2024gmt} or for more general kinematic configurations \cite{Aguilar:2024dlv}.

The general picture that emerges from continuum studies is that some of the form factors are always finite, as e.g. the form factor associated with the four-gluon tree level
tensor structure, while others seem to have a $\log$ divergence in the IR regime that is related to the the ghost being massless. 
For a discussion on tensor basis for the four-gluon Green functions see 
\cite{Binosi:2014kka,Gracey:2014ola,Eichmann:2015nra}. 
Within the limitations of the lattice approach, the lattice simulations seem to support such an outcome. The lattice QCD simulations and the Dyson-Schwinger studies 
provide similar results for the computed form factors.

The present proceeding is an update of the results reported in \cite{Colaco:2023qin,Colaco:2024gmt} for collinear kinematics, i.e. when all the four momenta are
proportional to each other, that considered $32^4$ and $48^4$ lattice sizes at $\beta = 6.0$. 
Herein, new data for the larger lattice volume generated with a larger gauge ensemble is discussed and the results of the two lattices are combined together.

\section{Lattice Setup and Definitions}

The simulations reported were performed with the Wilson action at $\beta = 6.0$. The corresponding lattice spacing, measured from the string
tension, is $a = 0.1016(25)$ fm or, equivalently, $1/a= 1.943(48)$ GeV. The gauge fields generated via importance sampling were rotated to the Landau gauge.
The computer simulations were performed using the Chroma \cite{Edwards:2004sx} and PFFT \cite{PFFT} libraries and run using 
the Navigator supercomputer at the University of Coimbra. Details on the gauge fixing algorithm and definitions of the gluon field in real and momentum spacetime
can be found in \cite{Colaco:2024gmt}.

The primary quantities accessed on a lattice simulation are the complete Green functions that for the four-gluon case is, in momentum spacetime, given by 
\begin{equation}
 \mathcal{G}^{a_1 \cdots a_4}_{\mu_1 \cdots \mu_4} (p_1, \, \cdots, \, p_4) = 
  \langle ~ A^{a_1}_{\mu_1} (p_1) ~ \cdots ~ A^{a_4}_{\mu_4} (p_4)~\rangle
  \label{Def:GreenFunctionN}
\end{equation}
and where $\langle ~ \rangle$ stands for ensemble average. The problem to access the corresponding 1PI Green function being that
$\mathcal{G}^{a_1 \cdots a_4}_{\mu_1 \cdots \mu_4} (p_1, \, \cdots, \, p_4)$ is a sum of disconnected parts, contributions that are proportional to
the three gluon 1PI functions and a term proportional to the four-gluon 1PI Green function that we are interested in measure, see the discussion in \cite{Colaco:2024gmt}. 
One way to suppress all the contributions but the latter one it by a proper choice of the momentum and taking profit that, in the Landau gauge, the gluon propagator
is transverse. The disconnected parts are products of gluon propagators and they can be removed from the complete Green function by choosing the momenta
of the external legs such that $p_i + p_j \ne 0$ for all $i$ and $j$.
On the other hand, due to the orthogonality of the gluon propagator, for the kinematical configurations such that
\begin{equation}
p_1 = p, \qquad p_2 = \eta p, \qquad p_3 = \lambda p, \qquad p_4 = -  ( 1 + \eta + \lambda ) p \ ,
\end{equation}
with $\eta$ and $\lambda$ being real numbers, the contributions proportional to the three gluon 1PI Green function are eliminated from the complete Green function. For
this family of kinematical configurations the complete Green function can be decomposed in terms of 1PI functions as
\begin{eqnarray}
& & 
\mathcal{G}^{abcd}_{\mu\nu\eta\zeta} (p_1, \, p_2, \, p_3, \, p_4) ~ = ~
\Big( P_\perp (p_1) \Big)_{\mu\mu^\prime} \, \Big( P_\perp (p_2) \Big)_{\nu\nu^\prime} \, 
\Big( P_\perp (p_3) \Big)_{\eta\eta^\prime} \, \Big( P_\perp (p_4) \Big)_{\zeta\zeta^\prime} 
\nonumber \\
& & \quad
D(p^2_1) \, D(p^2_2) \, D(p^2_3) \, D(p^2_4) \nonumber \\
& & \quad
\Bigg( F^{(0)}(p^2_1, \dots) \,\widetilde{\Gamma}^{(0)}  \,  ^{abcd}_{\mu^\prime\nu^\prime\eta^\prime\zeta^\prime}   
                    \nonumber \\
          & & \hspace{3cm}
          ~+~   
          F^{(1)}(p^2_1, \dots) \, \widetilde{\Gamma}^{(1)} \,  ^{abcd}_{\mu^\prime\nu^\prime\eta^\prime\zeta^\prime}   
          ~+~   
          F^{(2)}(p^2_1, \dots) \, \widetilde{\Gamma}^{(2)} \,  ^{abcd}_{\mu^\prime\nu^\prime\eta^\prime\zeta^\prime}   ~+  \cdots \Bigg)
          \label{G:FFdef}
\end{eqnarray}
where the gluon propagator reads
\begin{equation}
 \mathcal{G}^{a_1 a_2}_{\mu_1 \mu_2} (p,  \, p^\prime) = 
  \langle ~ A^{a_1}_{\mu_1} (p) ~ A^{a_2}_{\mu_2} (p^\prime)~\rangle =
  V \, \delta(p + p^\prime) \, \delta^{a_1 a_2} \left( \delta_{\mu_1 \mu_2} - \frac{p_{\mu_1} p_{\mu_2}}{p^2_1} \right) \, D(p^2) \ ,
\end{equation}
where the term in parenthesis in the r.h.s. is $\left( P_\perp (p) \right)_{\mu_1\mu_2}$. 
For a description of the gluon propagator $D(p^2)$ see e.g. \cite{Oliveira:2012eh,Duarte:2016iko} and references therein.
In  Eq. (\ref{G:FFdef}) the term in parenthesis is the tensor
decomposition of the four-gluon 1PI Green function. Herein we will consider that the Green function is completely described by three form factors that
are associated with the following tensor operators
\begin{eqnarray}
 \widetilde{\Gamma}^{(0)} \,  ^{abcd}_{\mu\nu\eta\zeta}  & = &
   f_{abr} f_{cdr} \Big( \delta_{\mu\eta} \, \delta_{\nu \zeta} - \delta_{\mu\zeta} \, \delta_{\nu \eta} \Big) ~ + ~
       f_{acr} f_{bdr} \Big( \delta_{\mu\nu} \, \delta_{\eta\zeta} - \delta_{\mu\zeta} \, \delta_{\nu \eta}  \Big) 
  \nonumber \\  & & \qquad
       + ~
       f_{adr} f_{bcr} \Big( \delta_{\mu\nu} \, \delta_{\eta\zeta}  - \delta_{\mu\eta} \, \delta_{\nu \zeta}  \Big) \ ,
       \label{Eq:four_glue_proj-tree}
   \\
 \widetilde{\Gamma}^{(1)} \,  ^{abcd}_{\mu\nu\eta\zeta}  & = &
   d_{abr} d_{cdr} \Big( \delta_{\mu\eta} \, \delta_{\nu \zeta} + \delta_{\mu\zeta} \, \delta_{\nu \eta} \Big) ~ + ~
       d_{acr} d_{bdr} \Big( \delta_{\mu\zeta} \, \delta_{\nu \eta} + \delta_{\mu\nu} \, \delta_{\eta\zeta} \Big)
  \nonumber \\  & & \qquad
        + ~
       d_{adr} d_{bcr} \Big( \delta_{\mu\nu} \, \delta_{\eta\zeta}  + \delta_{\mu\eta} \, \delta_{\nu \zeta}  \Big) \ ,
  \label{Eq:four_glue_proj-2}   
  \\
 \widetilde{\Gamma}^{(2)} \,  ^{abcd}_{\mu\nu\eta\zeta}  &  = &
\Big(  \delta^{ab}  \, \delta^{cd} + \delta^{ac}  \, \delta^{bd} + \delta^{ad}  \, \delta^{bc}  \Big) ~
\Big(  \delta_{\mu\nu}  \, \delta_{\eta\zeta} + \delta_{\mu\eta}  \, \delta_{\nu\zeta} + \delta_{\mu\zeta}  \, \delta_{\nu\eta} 
\Big) 
\label{Eq:four_glue_proj-3}
\end{eqnarray}
with $f_{abc}$ and $d_{abc}$ being the fully anti-symmetric and fully symmetric  SU(3) structure constants.
$\widetilde{\Gamma}^{(0)}$ is the tensor structure associated with four-gluon tree level Feynman rule.
As discussed in \cite{Colaco:2024gmt} it is possible to define projectors that allow to access each of the
 form factors $F^{(i)}$ from the complete Green function. The following kinematical configurations were chosen
\begin{displaymath}
  \begin{array}{c@{\hspace{0.75cm}}c@{\hspace{0.75cm}}c@{\hspace{0.75cm}}c@{\hspace{1cm}}l}
  \mathbf{p_1} & \mathbf{p_2} & \mathbf{p_3} & \mathbf{p_4} & \\
\\
  0     & p & p & -2 \, p & \mbox{referred as } (0, \, p, \, p, \, -2p)\\
  0     & p & 2 \,p & -3 \, p  & \mbox{referred as } (0, \, p, \, 2 p, \, -3p)\\
  p     & p & p & -3 \, p & \mbox{referred as } (p, \, p, \, p, \, -3 p) \\
  \end{array}
\end{displaymath}
In all cases an average over equivalent momenta is performed. The size of the gauge ensembles is about 9040.

\section{The Lattice Form Factors}

\begin{figure}[h] 
   \centering
   \includegraphics[width=2.92in]{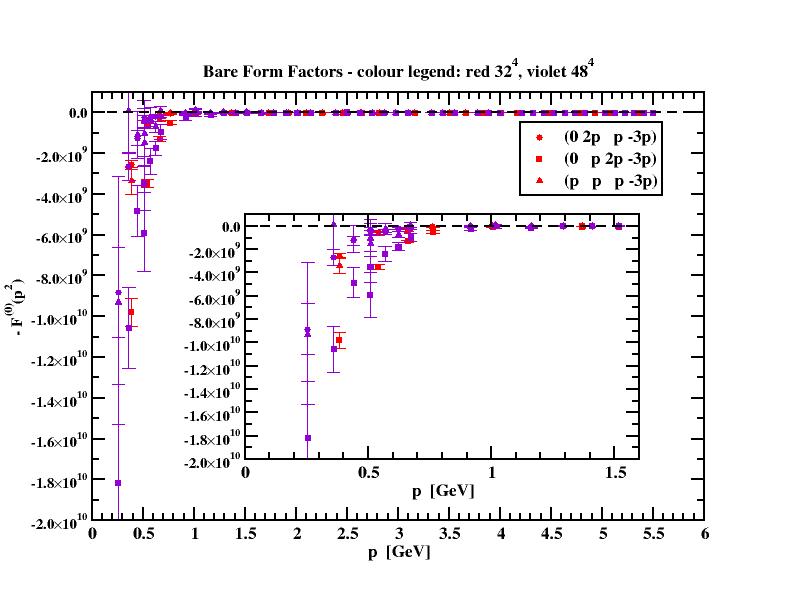} ~
   \includegraphics[width=2.92in]{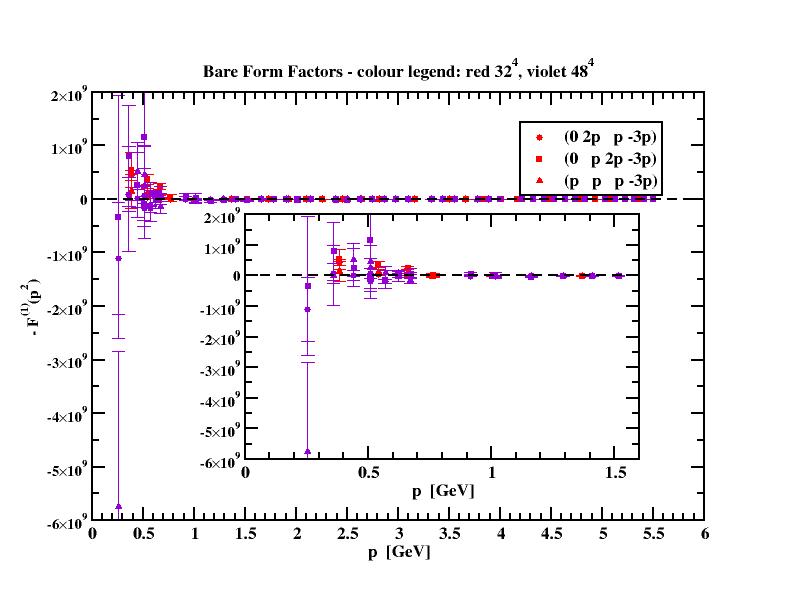} \\
   \includegraphics[width=2.92in]{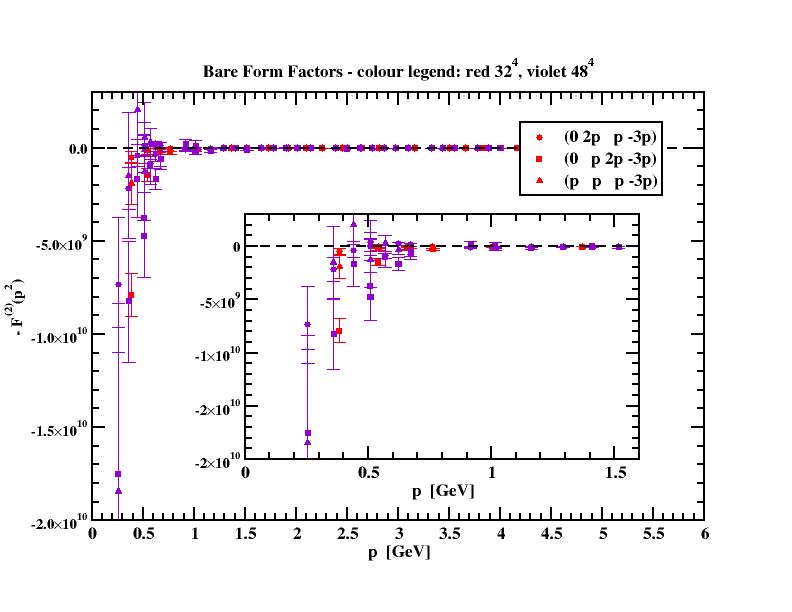} 
   \caption{Un-amputated form factors associated with the complete Green function.}
   \label{fig:1}
\end{figure}

In the following we report on the estimation for the various form factors. In Fig. \ref{fig:1} the un-amputated form factors associated with
the complete Green function, i.e. that include the contributions of the gluon propagator, is shown as a function of momentum $p$.
Within the statistical errors the results for the two volumes agree with each other, within each family of momenta. 
The difference between the data for the various types of momenta reported comes from the form factors being functions of the symmetric combination
of momenta $s = \sum_i p^2_i/4$, rather than functions of $p$.

\begin{figure}[h] 
   \centering
   \includegraphics[width=2.92in]{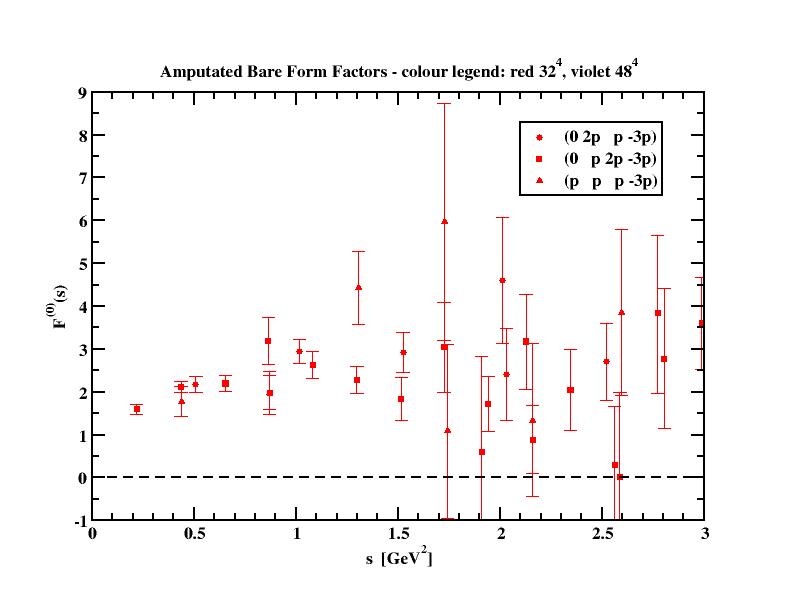} ~
   \includegraphics[width=2.92in]{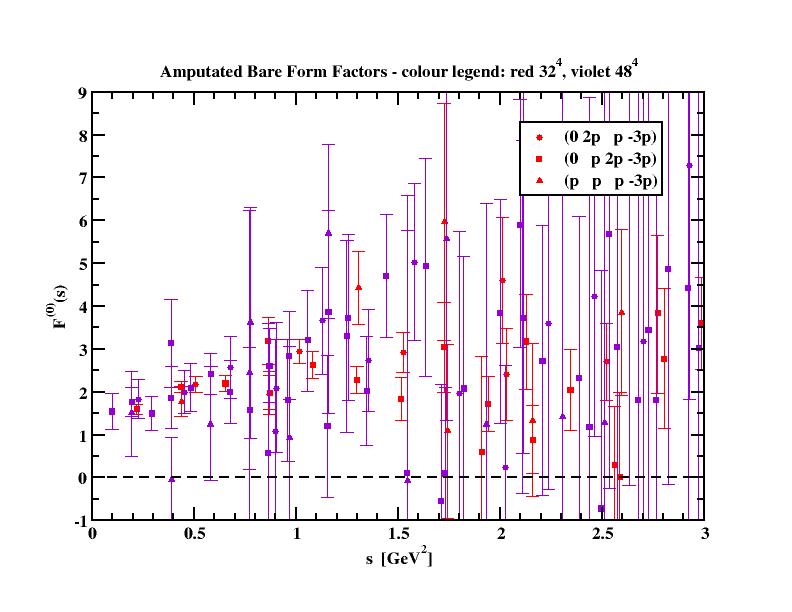}
   \caption{Amputated form factor associated with the 1PI Green function tree level tensor structure as a function of $s$. The left plot shows the data for the $32^4$ lattice, while
   the right plot shows the data for the two volumes used in the simulation.}
   \label{fig:2}
\end{figure}

\begin{figure}[h] 
   \centering
   \includegraphics[width=4in]{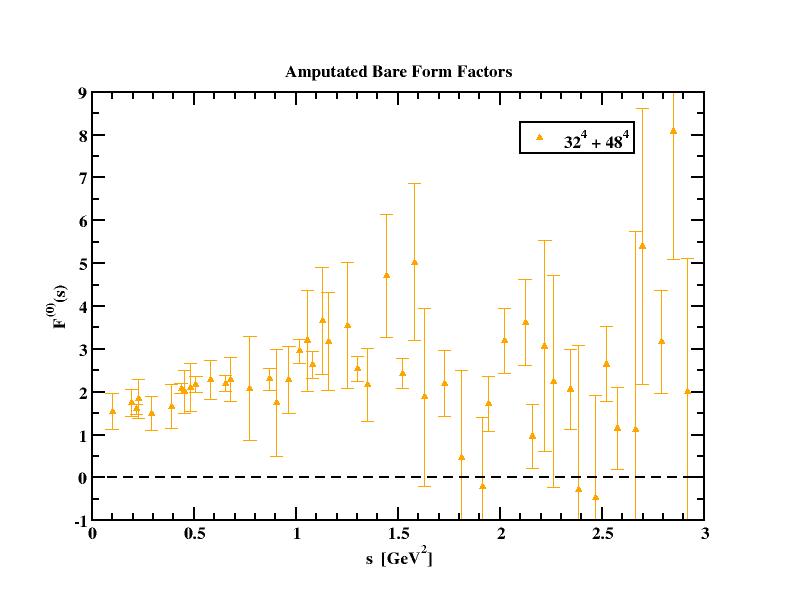}
   \caption{Amputated form factor $F^{(0)}(s)$ after averaging over momenta that differ by less than $2.5\%$ and taking into account the data of both lattice volumes.}
   \label{fig:3}
\end{figure}

\noindent
Indeed, in Fig. \ref{fig:2} the data for the amputated Green function after projection on $\widetilde{\Gamma}^{(0)}$, i.e. for $F^{(0)}(p^2)$, is shown as a function of $s$,
with the difference between the outcome of the various kinematical configurations is resolved. The right plot shows the data computed for the smaller lattice volume,
while the right plot shows the data for the larger volume. The form factors being functions of $s$ and not of $p$ is a manifestation of Bose symmetry. 
Another relevant point about the data in Fig. \ref{fig:2} is that, for the same $s$, there are various estimations of the form factors. This allow to combine this
data and compute a weighted average of these equivalent points. For the form factor associated with the tree level tensor operator, i.e. $\widetilde{\Gamma}^{(0)}$,
such an average is reported in Fig. \ref{fig:3}. Similar procedures can be built for the remaining form factors, i.e. for $F^{(1)}(s)$ and $F^{(2)}(s)$.

In what concerns $F^{(0)}(s)$ itself, the data in Fig. \ref{fig:3} is compatible with a constant form factor for $s \gtrsim 1$ GeV$^2$, while for smaller $s$ the lattice points
suggests a slight decrease in $F^{(0)}(s)$ as one approaches the deep infrared region. For $s \gtrsim 1$ GeV$^2$ the data is well described by a constant. This
was checked by fitting the data points. Looking only at those points with smaller statistical errors suggest this type of behaviour.

\begin{figure}[h] 
   \centering
   \includegraphics[width=2.92in]{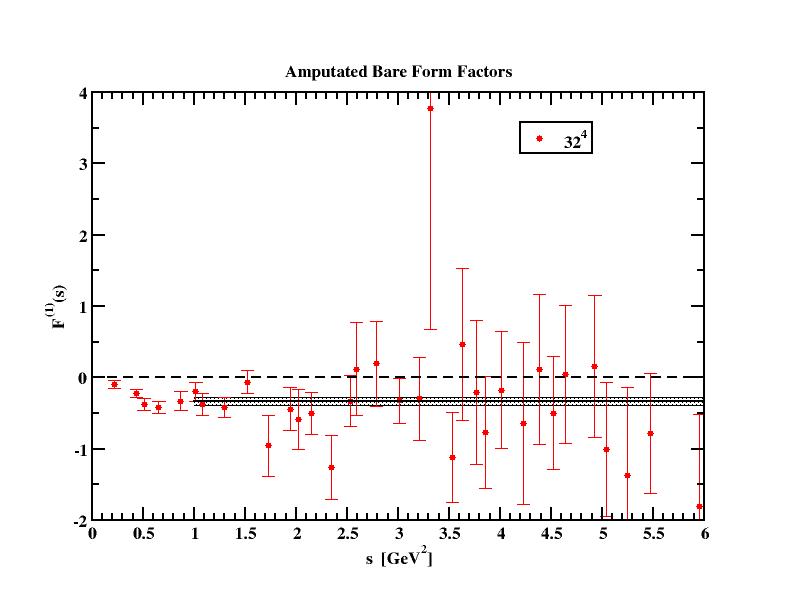} ~
   \includegraphics[width=2.92in]{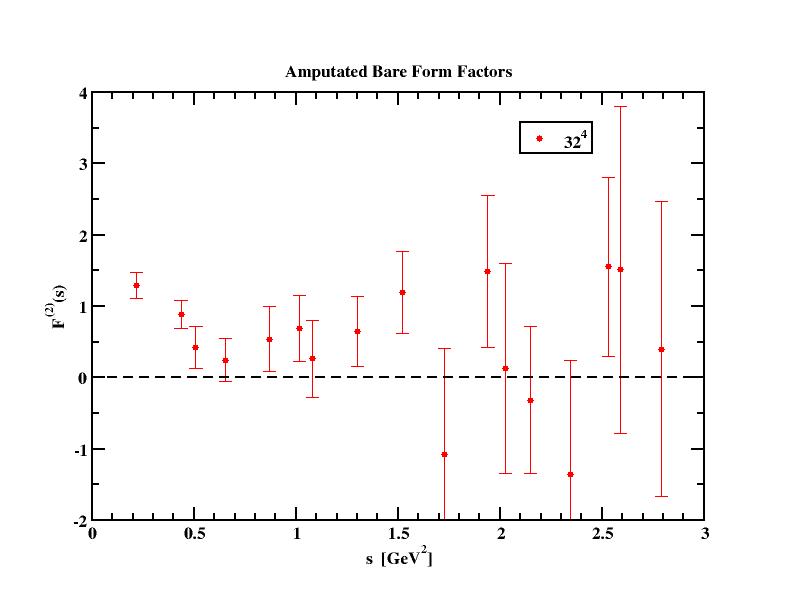}
   \caption{The form factor $F^{(1)}(s)$ (left) and $F^{(2)}(s)$ (right) after averaging over equivalent momenta using only the $32^4$ data.}
   \label{fig:4}
\end{figure}

The non-amputated versions of $F^{(1)}$ and $F^{(2)}$ are about one order of magnitude smaller when compared with $F^{(0)}$; see Fig. \ref{fig:1}. This also implies that the Monte
Carlo signal to noise ratio for these form factors is poorer. It follows that for the larger lattice, given that the size of the gauge ensembles is equivalent, the
relative statistical errors are much larger than for the smaller volume. No extra information on the form factors comes by considering the data from the $48^4$ lattice
for $F^{(1)}$ and $F^{(2)}$.
Then, when performing averages over equivalent $s$ we disregard the larger volume data. The lattice for $F^{(1)}(s)$ and $F^{(2)}(s)$ after this average uses
only the data from the $32^4$ lattice simulation and is reported in Fig. \ref{fig:4}.

The form factor $F^{(1)}(s)$ is compatible with a negative constant for $s \gtrsim 1$ GeV$^2$ and starts to increase as $s \rightarrow 0$. The lines in left plot of
Fig. \ref{fig:4} are the result of a $\chi^2/d.o.f.$ minimisation using a constant function.  Similar observations can be made for $F^{(2)}(s)$ that is reported on the right
plot of the same Fig. Note that for the later form factor only data for $s \leqslant 3$ GeV$^2$ is shown. For larger values of $s$ the estimated statistical errors become
rather large.

\section*{Acknowledgments}

\noindent
This work was partly supported by the FCT – Funda\c{c}\~ao para a Ci\^encia e a Tecnologia, I.P., under Projects Nos. 
UIDB/04564/2020 (\url{https://doi.org/10.54499/UIDB/04564/2020}), 
UIDP/04564/2020 (\url{https://doi.org/10.54499/UIDP/04564/2020})
and CERN/\-FIS-\-PAR\-/\-0023\-/2021. 
P. J. S. acknowledges financial support from FCT contract CEECIND/00488/2017 
(\url{https://doi.org/10.54499/CEECIND/00488/2017/CP1460/CT0030}).
The authors acknowledge the Laboratory for Advanced Computing at the University of Coimbra (\url{http://www.uc.pt/lca}) for providing 
access to the HPC resources that have contributed to the research within this paper. Access to Navigator was partly supported by the FCT Advanced Computing 
Project 2021.09759.CPCA (\url{https://doi.org/10.54499/2021.09759.CPCA}),  project 2022.15892.CPCA.A2 (\url{https://doi.org/10.54499/2022.15892.CPCA.A2}
and project 2023.10947.CPCA.A2 (\url{https://doi.org/10.54499/2023.10947.CPCA.A2}).


\end{document}